\documentclass[prl,aps,longbibliography,twocolumn,nofootinbib, superscriptaddress]{revtex4-2}
\usepackage{graphicx}

\usepackage[nointegrals]{wasysym} %
\usepackage[export]{adjustbox}
\usepackage{xcolor}
\usepackage{amsmath,amsfonts,amssymb,latexsym}
\usepackage{hhline}
\usepackage{bm}
\usepackage{verbatim}
\usepackage{enumitem}
\usepackage{mathrsfs}
\usepackage{slashed}
\usepackage{empheq}
\usepackage{soul}

\usepackage{hyperref}
\usepackage{braket}
\newcommand{\bfk}{\mathbf{k}}
\newcommand{\bfh}{\mathbf{h}}
\newcommand{\bfj}{\mathbf{j}}
\newcommand{\bfE}{\mathbf{E}}
\newcommand{\bfq}{\mathbf{q}}

\newcommand{\bea}{\begin{eqnarray}}
\newcommand{\eea}{\end{eqnarray}}
\newcommand{\I}{\text{i}}
\newcommand{\td}{\text{d}}
\newcommand{\bk}{\mathbf{k}}

\usepackage{xcolor}

\usepackage[symbol]{footmisc}

\renewcommand{\thefootnote}{\fnsymbol{footnote}}

\begin{document}

\title{Signatures of hidden octupolar order from nonlinear Hall effects}

\date{\today}

\begin{abstract}
Detecting symmetry-breaking hidden orders with conventional probes has been a long-standing challenge in the field of magnetism.
Higher-rank multipolar ordering -- anisotropic charge and magnetization distributions arising from a combination of spin-orbit coupling and crystalline environments -- is a quintessential example of such hidden orders, where new protocols of direct detection remain highly desirable.
In this work, we propose non-linear Hall effects as a novel probe for multipolar ordering in metallic systems. 
Taking inspiration from the family of Pr-based heavy-fermion compounds, Pr(Ti,V)$_2$Al$_{20}$, we formulate a minimal cubic-lattice model of conduction electrons coupled to a ferro-octupolar order parameter.
The time-reversal-breaking order leads to a band structure that supports strong quadrupolar moments of the Berry curvature (BC).
Using a semi-classical Boltzmann formalism in conjunction with a symmetry analysis, we demonstrate that the BC quadrupoles produce a third harmonic generation of the Hall voltage $[V_H(3 \omega)]$ measurable in an AC Hall experiment.
Properties of the Hall response such as its anisotropy, its dissipationlessness, and its dependence on the order parameter are also examined. 
Our work encourages a new realm of investigation of multipolar ordering from non-linear transport experiments.
\end{abstract}

\author{Sopheak Sorn$^*$}
\affiliation{Institute of Quantum Materials and Technology, Karlsruhe Institute of Technology, 76131, Karlsruhe, Germany}	
\thanks{Authors contributed equally to this work.}
\author{Adarsh S. Patri$^*$}
\affiliation{Department of Physics, Massachusetts Institute of Technology, Massachusetts 02139, USA}	
\thanks{Authors contributed equally to this work.}
	\maketitle

\def\thefootnote{*}\footnotetext{Authors contributed equally to this work.\\
Email: sopheak.sorn@mail.utoronto.ca\\
Email: apatri@mit.edu}\def\thefootnote{\arabic{footnote}}

A deep understanding of quantum phases of matter ultimately requires experimental probes of the low-energy degrees of freedom. 
The celebrated Hall effect is an archetypal example supporting this philosophy, from the remarkable insight it 
can provide of the underlying properties of a system, ranging from the electron-hole nature of the charge carriers to the topology of the electronic band structure \cite{TKNN, qhe_review, ahe_review, qah_review, she_review}.
At the core of the Hall effects is the role of the BC of the occupied electronic bands.
Semiclassically, for a spin-orbit-coupled system featuring a spin-imbalanced occupation number associated with a net magnetization, BC acts as a spin-dependent, momentum-space magnetic field to transversely deflect the electrons, giving rise to a well-defined Hall response \cite{kl_he, niu_semiclassical, niu_sundaram}.
Occurring even in the absence of a magnetic field, this anomalous Hall effect (AHE) scales with the magnetization, and can shed light on the canonical, ferromagnetically ordered ground state. 
This brings forth the intriguing question of whether Hall effect (or its generalizations) is flexible enough to provide discriminating signatures of unconventional broken-symmetry phases of matter.

Recently, non-linear Hall effects, where the transverse Hall voltage scales non-linearly with the applied charge current, has gained significant attention \cite{Fu2015, Moore2016, Ma2019, firstBCM, BCP0, Lai2021, Wang2021, Liu2021, nlhe_review, Nagaosa2022, Nagaosa2022b, Law2023}. Non-linear Hall effect can be understood as a natural multipolar extension of the mathematical AHE framework, where instead of merely considering the Brillouin-zone integral of the BC, one considers BC dipoles, BC quadrupoles, and beyond \cite{Fu2015, Moore2016, firstBCM, nlhe_review, Law2023}.
Due to the higher-rank nature of these BC multipolar moments, the resulting non-linear Hall effects can arise even without a net magnetization.
Indeed, evidence of BC-dipole-driven non-linear Hall effect has been seen in a variety of non-magnetic compounds \cite{Ma2019, Kang2019, Hu2022, Huang2022, nlhe_review} and in non-centrosymmetric compensated anti-ferromagnets \cite{Shao2020}
Similarly, there has been evidence of BC-quadrupole-driven non-linear Hall effects in altermagnets \cite{fang2023}, an emerging family of unconventional compensated collinear anti-ferromangets \cite{Sinova2022, Sinova2022_2}.

This lack of magnetization is reminiscent of a class of $d$ and $f$ electronic systems, where localized electronic wavefunctions support higher-rank multipolar moments \cite{multupole_rev_1, multupole_rev_2, multupole_rev_3}. These anisotropic charge and magnetization densities (characterized by quadrupolar, octupolar and higher-rank degrees of freedom) have been examined in a variety of contexts from actinide oxides (such as NpO$_2$ \cite{npo2_ref}) to $f$-electron heavy fermion systems \cite{review_exotic_multipolar,hfm_superconductivity_v,ye2023measurement, Chandra_2002, Tripathi_2007, Santander_2009, Kotliar_Haule_2009, Okazaki_2011, rau_2012} 
(including a system we will focus on here, Pr(Ti,V)$_2$Al$_{20}$) to pyrochlore quantum spin ices (Ce$_2$(Sn,Zr)$_2$O$_7$ \cite{sibille_2015, pengcheng_do_2019, gaulin_2019, fennell_2020, gaulin_2022} and Pr$_2$(Hf,Sn,Zr)$_2$O$_7$ \cite{phfo_1,phfo_2, phfo_nat_phys, pso_1, pso_2, pso_3, kimura_nakatsuji_ncomm_pzo_2013}) and to $d$-electron compounds including osmates and rhenates 
\cite{Chen2010, Chen2011, Fu2015_2, Harter2017, Lu2017, Hayami2018, Hirai2019, Maharaj2020, Paramekanti2020, Voleti2020, Lovesey2020, Svoboda2021, Lovesey2021, Pourovskii2021, Khaliullin2021, Churchill2022}.
Indeed, the ordering of higher-rank moments are notoriously difficult to directly detect with conventional local probes of magnetism, and have been appropriately placed under the umbrella of ``hidden orders" \cite{hidden_order_review}.
Despite recent proposals to employ lattice-based protocols to detect multipole-based phenomena (such as magnetostriction, ultrasound, and impurity-induced strains \cite{patri_unveil_2019, fisher_quadrupole, patri_do,patri_pzo, simon_2022, Voleti2023}), it is still highly desirable to devise new probes in order to shine light on these phases of matter. 
With the aforementioned promising successes of Hall measurements in systems lacking a net magnetization, we are primed to examine the detection of multipolar orders within the framework of \textit{non-linear} Hall effects.

\begin{figure}[t]
\includegraphics[width = 0.4\textwidth]{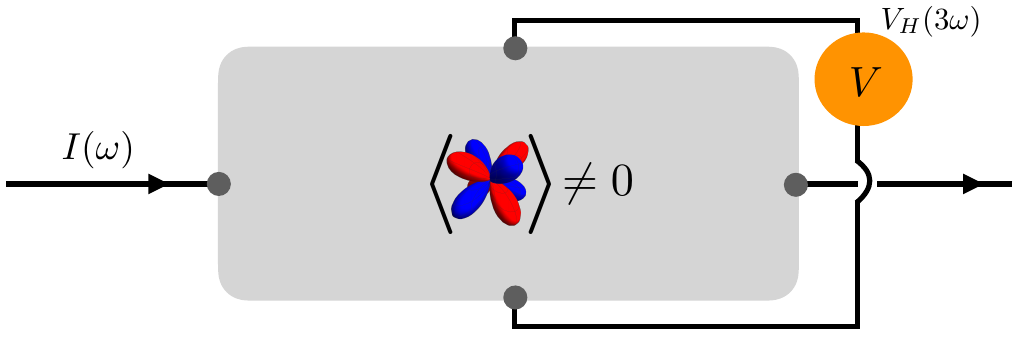} \\
\caption{Schematic of a Hall-bar
configuration to realize the non-linear Hall effect with an underlying octupolar order, $\phi \neq 0$. The third harmonic Hall response $\sigma^{(3\omega), H}_{abcd}$ is measured from the Hall voltage $V_H(3\omega)$ induced by an AC current $I(\omega)$ with a frequency $\omega$.} 
\label{fig_schematic_nlh}
\end{figure}

In this Letter, we explicitly demonstrate that metallic systems featuring a long-range ferro-octupolar order can be probed from a third-order Hall response. 
Taking a minimal cubic-lattice $e_g$ electron model coupled to a ferro-octupolar order parameter, we show that the onset of the order opens gaps at various band crossing points and induces strong BC quadrupolar moments.
The latter produces a third-harmonic generation detectable within a third-order Hall effect measurement (Fig. \ref{fig_schematic_nlh}).

\textit{Minimal Model of Octupolar Order.---}
In the family of Pr-based rare-earth compounds, Pr(Ti,V)$_2$Al$_{20}$, the combination of strong spin-orbit coupling and crystal-electric-field effects leads the Pr$^{3+}$ ions to hosting higher-rank multipolar moments:
(1) time-reversal-even quadrupolar moments $\mathcal{O}_{20} = \frac{1}{2}(3J_{z}^2 - J^2)$ and $\mathcal{O}_{22} = \frac{\sqrt{3}}{2}(J_{x}^2 - J_{y}^2)$, and (2) a time-reversal-odd octupolar moment $\mathcal{T}_{xyz} = \frac{\sqrt{15}}{6}\overline{J_x J_y J_z}$; the overline indicates a symmetrized operator product over the angular-momentum Stevens operators.
Via the conduction-electron-mediated RKKY-like interaction between the localized moments, long-range ordering is permitted to develop.
We consider the simple scenario of $\bfq = 0$ ferro-octupolar order given by the spatially uniform order parameter $\phi = \langle \mathcal{T}_{xyz}\rangle$.

Recent de-Haas van Alphen studies on PrTi$_{2}$Al$_{20}$\cite{dhva_nagashima_2014, dhva_nagashima_2020}  have indicated a well-localized Fermi surface about the zone center.
As such, the low-energy conduction electrons can be characterized in terms of irreducible representations of the corresponding $O_h$ point group symmetry. 
In this work, we consider the following Slater-Koster tight-binding model of $e_g$ electrons on a cubic lattice, which is inspired by the Pr-based compounds,
\begin{align}
\mathcal{H}_0(\bfk)= h_0(\bfk)\tau^0 + h_1(\bfk) \tau^x + h_3(\bfk) \tau^z,
\end{align}
where $\tau$'s are Pauli matrices in the two-dimensional orbital space $\{d_{x^2 - y^2}, d_{3z^2 -r^2} \}$. $h_s(\bfk)$ are the momentum-dependent form factors whose full expressions are given in the Supplementary Materials (SMs). Near the zone center, their quadratic-order expansions are given by simple expressions: $h_0 (\mathbf{k}) = \text{constant} + \tilde{\lambda}k^2$, $h_1(\bfk) = \frac{\sqrt{3}\lambda}{2} \left(k_x^2 - k_y^2 \right)$ and $h_3(\bfk) = \frac{\lambda}{2} \left(3 k_z^2 - k^2 \right)$, where $\lambda \text{ and } \tilde{\lambda} $ are associated with the overlap integrals (see SMs).

Equipped with these orbital and spin degrees of freedom, the conduction electrons couple to the overlaying ferro-octupolar order via the symmetry-allowed term \cite{patri_kondo_2020, patri_kondo_2021}
, $\mathcal{H}_{\phi}(\bfk) = \phi \tau^y,$
where we have absorbed the coupling constant into the octupolar order parameter.
We note that the spin plays the role of a spectator degree of freedom and is included as an implicit identity matrix in both the kinetic and coupling terms \cite{patri_kondo_2020, patri_kondo_2021}. 
Therefore, the full model is $\mathcal{H}_0 + \mathcal{H}_{\phi} = h_0(\bfk) \tau^0 + \bfh(\bfk)\cdot \boldsymbol{\tau}$, where $\bfh=(h_1, h_2, h_3)^T$. The band eigenvalues are $\varepsilon_{\bfk, \pm} = h_0(\bfk) \pm |\bfh(\bfk)|$. 

\begin{figure}[t]
    \includegraphics[width=0.4\textwidth]{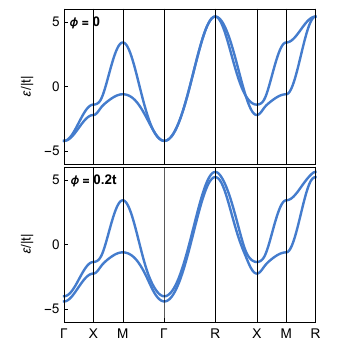}
\caption{Cubic-lattice electronic band structure in the absence (presence) of octupolar ordering $\phi$.
The octupolar ordering introduces a splitting of the degenerate band structure. The following overlap integrals, defined in SMs, are used here and for the numerical results: ($t$, $u$, $v$, $w$) = ($-1$, $-0.6$, $-0.1$, $0.2$).
}
\label{fig_bandstructure_berry}
\end{figure}

Figure \ref{fig_bandstructure_berry} shows representative band structures in the absence and presence of the octupolar ordering.
Notably when $\phi = 0$, there are many degenerate points such as those on the $\Gamma M$ line and $RX$ line. These degenerate points are protected by the mirror symmetry of the mirror planes, $(110), (101), (011), (\bar{1}10),(\bar{1}01), \text{and } (0\bar{1}1)$. On the other hand, the entire degenerate $\Gamma R$ line is protected by the mirrors \emph{and} the thee-fold rotation around the [111] axis;
see SMs for a detailed description. A non-zero $\phi$ breaks these symmetries and lifts the degeneracy of the bands, leading to substantial BC near the gaps.

\textit{Linear and non-linear Hall effects.---} 
When $\phi \neq 0$, the magnetic point group of the system is $m\bar{3}m'$. Correspondingly, the linear-response conductivity tensor $\sigma_{ab}$ is proportional to a 3-by-3 identity matrix in the Cartesian basis directions. 
The two-fold rotational symmetries about the three coordinate axes set the off-diagonal elements to zero, while the diagonal elements are identical due to the three-fold rotation around the [111] axis.
Therefore, from these symmetry considerations, the anomalous Hall effect is absent; from linear response theory, one can see this explicitly from the vanishing Brillouin-zone integral of the BC, $\sum_n\int \td \bfk f_0(\varepsilon_{\bfk n})\Omega^n_c(\bfk)$ = 0, where $n$ is the band index, and the local BC is $\boldsymbol{\Omega}^n(\bfk) = \boldsymbol{\nabla}_{\bfk} \times \I \bra{\bfk n} \boldsymbol{\nabla}_{\bfk} \ket{\bfk n}$.
A similar analysis (as described in SMs) shows that the the second-order Hall response is also zero due to the inversion symmetry.

The Hall physics of our system is thus dominated by the third-order response, where a third-order current, $j^{(3)}_a(t) = \text{Re}\left[j^{(3)}_a(\omega) e^{\I \omega t}\right]$, is induced by an applied electric field $E_a(t) = \text{Re}\left[E_a(\omega)e^{\I \omega t} \right]$. 
With a Hall-bar-like experimental setup in mind (Fig. \ref{fig_schematic_nlh}), we restrict ourselves to linearly polarized electric field, where $E_a(\omega)$ can be made real-valued by redefining the initial time. 
The induced current can be expressed in terms of a third-order response function, $j^{(3)}_a(2\omega \pm \omega) = \sigma_{abcd}^{(2\omega \pm \omega)} E_b(\omega)E_c(\omega)E_d(\pm \omega)$. 
There are two types of the currents with frequencies $\omega$ and $3\omega$. 
The rest of the main text will focus on the third-harmonic $\sigma_{abcd} \equiv \sigma_{abcd}^{(3\omega)}$, while a straightforward generalization to $\sigma_{abcd}^{(\omega)}$ is discussed in SMs. 
Using a semi-classical Boltzmann formalism, it can be shown that $\sigma_{abcd}$ consists of two parts: $\sigma_{abcd} = \sigma_{abcd}^D + \sigma_{abcd}^H$, where $\sigma_{abcd}^D$ is the time-reversal-even Drude-like part, and $\sigma_{abcd}^H$ is the time-reversal-odd Hall-like part \cite{Fu2015, Moore2016, firstBCM, Law2023},
\begin{align}
\sigma^D_{abcd} &= \frac{e^4}{4\hbar} \sum_n\int \td\bfk \frac{f_0(\varepsilon_{\bfk n}) \partial_a\partial_b\partial_c\partial_d \varepsilon_{\bfk n}}{(\hbar \widetilde{\omega})(\hbar \widetilde{2\omega})(\hbar \widetilde{3\omega})},\\
\sigma^H_{abcd} &= - \frac{e^4}{12\hbar} \frac{\epsilon_{hab}Q_{cdh} + \epsilon_{hac}Q_{dbh} + \epsilon_{had}Q_{bch}}{(\hbar \widetilde{\omega})(\hbar \widetilde{2\omega})},
\label{eq:nonlinearHall}
\end{align}
$\widetilde{m \omega} \equiv \I m\omega + 1/\tau$, $m$ is an integer, and $\tau$ is the relaxation time in the Boltzmann formalism. 
$Q_{abc}^n$ is the BC quadrupolar moment for a given band, which is manifestly odd under time reversal,
\begin{align}
    Q_{abc} & \equiv \sum_n Q_{abc}^n = \sum_n \int \td \bfk f_0(\varepsilon_{\bfk n}) \partial_{k_a}\partial_{k_b} \Omega_c^n(\bfk).
    \label{eq:BCQ}
\end{align}
Similarly, the induced current can be split into a Drude-like and a Hall-like part. The Levi-Civita symbol in Eq. \ref{eq:nonlinearHall} leads to an orthogonality between the Hall current and the applied field, so the Joule heating is absent. 
The rest of the Letter will focus on the time-reversal-odd dissipationless response. We note that we have ignored a third-order nonlinear Hall contribution from a correction to the Berry connection due to the electric field  since it is time-reversal even and can exist even without the octupolar order \cite{BCP1,BCP2,BCP3}; see SMs for further discussions.

The $m\bar{3}m'$ magnetic point group ensures that there is only one independent component of $\sigma_{abcd}^H$ whose value is denoted by $\sigma_H$,
\begin{align}
    \sigma_{xxyy}^H = \sigma_{yyzz}^H = \sigma_{zzxx}^H = \sigma_H,\nonumber\\
    \sigma_{xxzz}^H = \sigma_{zzyy}^H = \sigma_{yyzz}^H = - \sigma_H,
    \label{eq:Hallcomponent}
\end{align}
where $\sigma_H \propto Q_{xyz}$. For brevity, the components obtained from permuting the last three indices are not shown, while the rest of $\sigma^H_{abcd}$ components are zero. 

\begin{figure}
    \center
    \includegraphics[width=0.35\textwidth]{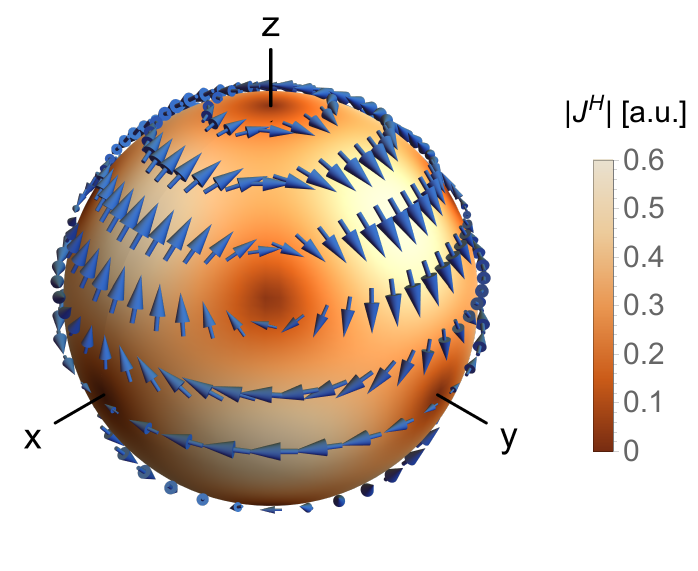}
    \caption{Schematic illustration of the anisotropic nature of the non-linear Hall response: at each point on $S^2$, the arrow lying in the tangent plane represents the induced Hall current $\bfj^H$ in response to an applied electric field (that is normal to the surface). The color map (and the size of the arrows) illustrates the magnitude of $\bfj^H$, which vanishes at fourteen high-symmetry points.}
    \label{fig:anisotropy}
\end{figure}

\textit{Anisotropic nature of the induced Hall current.---} 
Using a spherical-coordinate representation of $\bfE(\omega) = |\bfE(\omega)| \hat{E}$, where $\hat{E} = 
(\cos \varphi \sin \theta, \sin \varphi \sin \theta, \cos \theta)^T$, the induced Hall current is given by
\begin{align}
    \bfj^H &\sim \begin{pmatrix}
        \sin \theta \cos \varphi (\sin^2 \theta \sin^2 \varphi - \cos^2 \theta)\\
        \sin \theta \sin \varphi (\cos^2 \theta - \sin^2 \theta \cos^2 \varphi)\\
        \cos \theta \sin^2 \varphi (\cos^2 \varphi - \sin^2 \varphi)
    \end{pmatrix}.
\end{align} 
If we define $\hat{E}$ by a point on an $S^2$ sphere, $\bfj^H$ can be represented by a tangent vector at that point. In this manner, Figure \ref{fig:anisotropy} illustrates the profile of $\bfj^H$ (blue arrows) featuring a strong dependence on the orientation of $\hat{E}$. 
The magnitude of $\bfj^H$ is given by the color map of the inner sphere. 
$\bfj^H$ vanishes at fourteen points where $\hat{E}$ lies on the high-symmetry axes of [100], [010], [001], [111], [$\Bar{1}$11], [1$\Bar{1}$1], and [$\Bar{11}1$].

\textit{Berry curvature quadrupole.---} 
The magnitude of the Hall response is proportional to the BC quadrupole moments. In our model, BC is given by
\begin{align}
    \Omega^{\pm}_a(\bfk) &= \mp \frac{\epsilon_{abc}}{4} \hat{h}\cdot \partial_{k_b} \hat{h} \times \partial_{k_c} \hat{h},
\end{align}
where $\hat{h}(\bfk) = \bfh(\bfk)/|\bfh(\bfk)|$.
$\Omega^{\pm}_a(\bfk)$ is non-zero only when the vector field $\hat{h}(\bfk)$ is locally non-coplanar. 
In the absence of the octupolar order, $\hat{h}(\bfk)$ is co-planar, so BC vanishes everywhere. 
To provide intuition for this argument, it is instructive to consider the low-density limit, where the BC for the $\pm$ bands near the zone center is given by,
\begin{align}
\boldsymbol{\Omega}^\pm (\mathbf{k}) = \pm \frac{\sqrt{3} \lambda^2 \phi}{ (h_1^2 + h_3^2 + \phi^2)^{3/2}} \begin{pmatrix} k_y k_z, k_x k_z, k_x k_y \end{pmatrix}^T .
\label{eq_bc_low_density}
\end{align}
We can clearly see that in the absence of the time-reversal-breaking octupolar order, BC vanishes as expected (as inversion is a symmetry in this cubic system).
The corresponding BC quadrupole is given by,
\begin{align}
\label{eq_bc_quad_low_den}
 {Q_{xyz}^{\pm}} = \pm {\sqrt{3} \lambda^2} \int \td \bfk \frac{  F_1(\mathbf{k}) \phi  +  F_3(\mathbf{k}) \phi^3 - \phi^5  }{ (h_1^2 + h_3^2 + \phi^2 )^{7/2}}.
\end{align}
$F_{1,3}$ are polynomial functions of momenta (see SMs). 

\begin{figure}[t]
    \centering
    \includegraphics[width=0.465\textwidth]{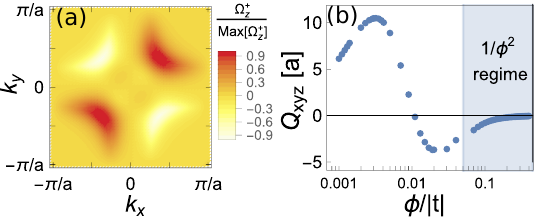}
    \caption{(a) Reciprocal-space distribution of the Berry curvature, $\Omega_z^+(\bfk)$, in the plane of $k_z = 0$, featuring a quadrupolar structure, which leads to a non-trivial Berry curvature quadrupole $Q_{xyz}$ and a non-zero third-order Hall response. (b) Non-monotonic dependence of $Q_{xyz}$ on the octupolar order parameter $\phi$. $Q_{xyz}$ is expressed in the unit of the cubic-lattice constant, $a$. Note that the x-axis is in a logarithmic scale. In the large-$|\phi|$ regime, $Q_{xyz}$ scales as $1/\phi^2$, as explained in the main text. }
    \label{fig:bcq}
\end{figure}

Figure \ref{fig:bcq}(a) shows the reciprocal-space profile of $\Omega_z^+(\bfk) = - \Omega_z^-(\bfk)$ when $\phi/|t| = 0.2$ within the plane of $k_z=0$. 
The BC profile possesses an azimuthal-angle dependence that is quadrupolar in nature thus hinting at the BC quadrupole, $Q_{xyz}$.
It is also apparent that there are BC hot spots with large BC concentration occurring near the gaped-out band crossing points along the $\Gamma M$ line (as well as $\Gamma R$ and $RX$ lines) due to the octupolar order.

Figure \ref{fig:bcq}(b) shows the dependence of $Q_{xyz}$ on $\phi$ at a fixed, low electron density of $0.004/a^3$. As will be clear, the low density is chosen since it permits a partly tractable analysis of the $\phi$ dependence of $Q_{xyz}$; for numerical results at other densities, see SMs. Our main observation is that the strength of the third-order Hall response is a non-monotonic function of the order parameter $\phi$. 
This is rather unusual compared with, for instance, the linear Hall response of a ferromagnet, where it generally increases with the magnetization order parameter\footnote{Deviations can arise when there are additional features in the systems, e.g. Weyl points in the band structures, or an interplay between intrinsic and extrinsic AHE, or topological skyrmion spin textures. These can lead to a non-monotonic relation between AHE and the magnetization. See e.g. \cite{ahe_review, Haham2011, Chen2013, Bartram2020, Lee2009, Neubauer2009} }. 

To provide intuition it is once again instructive to return to the low-density limit.
In the regime where $|\phi| \gg |t|$, we have two well-separated bands, and the electrons only occupy the lower band.
Appealing to Eq. \ref{eq_bc_low_density} and \ref{eq_bc_quad_low_den}, where the BC reduces to $\Omega_z^- (\mathbf{k}) = \frac{-\sqrt{3} \lambda^2 \text{sgn}(\phi)}{\phi^2} k_xk_y $, we find that the BC quadrupole reduces to $Q_{xyz} \sim \frac{\text{sgn}(\phi)}{\phi^2}$. 
This is in agreement with our numerical finding in Fig. \ref{fig:bcq}(b). This regime ends when the upper band also becomes occupied as $|\phi|$ is lowered. It can be shown that, after and near the Lifshitz transition, the function $Q_{xyz}(\phi)$ no longer scales like $1/\phi^2$ and becomes more complicated, as also evident from the numerical results (see SMs for a discussion.)

\textit{Signatures in AC Hall experiment.---} A suitable way to measure the non-linear Hall response due to the octupolar order is to perform a low-frequency\footnote{The low frequency is suited with the Boltzmann formalism since the inter-band transitions are suppressed; see Ref.\cite{nlhe_review} for a discussion.} AC Hall experiment, where an AC current of a frequency $\omega$ is sent across a Hall bar, and the Hall voltage in the transverse direction at a frequency of $3\omega$ is measured (see Fig. \ref{fig_schematic_nlh}). 
To remove the Drude-like contribution from the third-order response, as in Eq. \ref{eq:nonlinearHall}, the measured values need to be antisymmetrized between the two uniformly-polarized-domain measurements with the opposite order parameters, $\pm\phi$. 
A uniformly polarized octupolar domain could be procured, for example, by applying a magnetic field along the [111] direction \cite{patri_unveil_2019}.
The I-V relation between the amplitude of the applied current and that of the Hall voltage is expected to follow a cubic relation, $I \sim V^3$. 
Such a measurement is rather standard and has been used to measure various second-order Hall responses \cite{nlhe_review} and recently a third-order Hall response in FeSn \cite{sankar2023}.
For a future experimental design, it is worth-emphasizing that the Hall response depends strongly on the orientation of the applied current relative to the crystallographic directions, as demonstrated by Fig. \ref{fig:anisotropy}. 
Our results show that an appreciable BC quadrupole is possible in some parameter regime with $Q_{xyz} \sim 10^0- 10^2$\AA (see Fig. \ref{fig:bcq}(b) and assuming that the lattice constant $a \approx 14$ \AA for PrTi$_2$Al$_{20}$ \cite{Taniguchi_2016}.)
Remarkably, this is of a comparable strength to that of the BC quadrupole in FeSn, where the third-order Hall effect has been measured \cite{sankar2023}.

\textit{Impacts of possibly coexisting multipolar orders.---} 
It is possible, yet not necessary, to have the octupolar order coexisting with other multipolar orders, e.g. the quadrupolar order, $\tilde{\phi} = \langle \mathcal{O}_{22}\rangle$. $\tilde{\phi}$ alone does not break time-reversal symmetry, so the third-order Hall effect cannot be present with merely quadrupolar order. We have shown in SM that, in the coexistence case starting with a non-zero $\phi$, a growing $\tilde{\phi}$ increasingly suppresses BC quadrupoles. This outcome is attributed to the smearing effect of $\tilde{\phi}$ on the BC distribution. 
This can also be understood from the large-$\tilde{\phi}$ limit, as $\tilde{\phi}$ opens gaps at symmetry-protected band-crossing points, but it does not produce any BC hot spots (see SM); a nonzero $\phi$ no longer generates BC hot spots like before, hence a smoother BC distribution and smaller BC quadrupoles. Therefore, the third-order Hall response is the largest when $\phi$ stands alone.

\textit{Discussion.---} 
In this work, we demonstrated that a metallic system with ferro-octupolar order can exhibit a dissipationless third-order time-reversal-odd Hall response as a leading Hall phenomenon. This provides a new means to detect the onset of octupolar order.
We emphasized that this Hall effect arises despite the lack of a dipole moment, and we highlighted the key role of the Berry curvature quadrupole and its highly non-trivial dependence on the octupolar order parameter.
Our work encourages experimental investigations of multipolar ordering using non-linear Hall measurements (and more generally non-linear transport studies \cite{Nagaosa2020, Nagaosa2022}), in particular in the Pr-based systems.
Future theoretical studies are required to incorporate the effects of impurities, order parameter fluctuations, as well as influence of Kondo-like effects \cite{PhysRevB.98.235143,patri_kondo_2020, patri_kondo_2021, PhysRevResearch.3.013189} on non-linear Hall response.

\section{Acknowledgements}
We would like to thank Arun Paramekanti and Markus Garst for very helpful discussions. S.S. is supported by the Deutsche Forschungsgemeinschaft
through TRR 288 - 422213477 (project A11).
A.S.P. is supported by a Simons Investigator Award from the Simons Foundation.

\appendix
\section{Slater-Koster tight-binding model}

As detailed in the main text, we consider a kinetic energy model of $e_g$ electrons hopping on a simple cubic lattice.
In momentum space, we employ the basis representation $\psi_{\bfk a\sigma} = \left(d_{\bfk 1\sigma}, d_{\bfk 2 \sigma}\right)^T$, where $d_1 = d_{x^2-y^2}$ and $d_2 = d_{3z^2-r^2}$.
The corresponding kinetic term is $H_0 = \sum_{\bfk} \psi^{\dagger}_{\bfk a \sigma} \mathcal{H}^0_{ab}(\bfk) \psi_{\bfk b\sigma}$, where
\begin{align}
\mathcal{H}^0_{ab}(\bfk) = h_0(\bfk)\tau^0_{ab} + h_3(\bfk) \tau^3_{ab} + h_1(\bfk) \tau^1_{ab}. 
\end{align}
The $\tau$ matrices are Pauli matrices in the two-dimensional orbital space $\{d_1, d_2 \}$.
The forms of the three scalar functions, $h_0$, $h_1$ and $h_3$, involve the Slater-Koster hopping integrals,
\begin{align}
    h_0(\bk) = &(t+u) (c_x + c_y + c_z) \nonumber \\
    & + 2 (v+w) (c_x c_y + c_y c_z + c_z c_x)  \\
    & + (4 z) c_x c_y c_z \nonumber \\
    h_1(\bk) = & \frac{\sqrt{3}}{2} (c_x - c_y)\left[2c_z (v-w) + (u-t)\right],  \\  
    h_3(\bk) = & \frac{1}{2}(t-u)(c_x + c_y - 2 c_z) \nonumber \\
    & + (v-w)(c_x c_z + c_y c_z - 2 c_x c_y),
\end{align}
where $c_{i} = \cos(k_i a)$, $s_i = \sin(k_ia)$, and $t,u,v,w, z$ are the hopping integrals, which, in the Slater-Koster notation, correspond to the overlap integrals involving first, second, and third nearest neighbor bonds $E_{3z^2-r^2, 3z^2-r^2}(001)$, $E_{x^2-y^2, x^2-y^2}(001)$, $E_{3z^2-r^2, 3z^2-r^2}(110)$, $E_{x^2-y^2, x^2-y^2}(110)$, and $E_{x^2-y^2, x^2-y^2}(111)$, respectively. $h_1(\bk)$ corresponds to the inter-orbital hybridization between the $x^2-y^2$ and the $3z^2-r^2$ orbitals.
We neglect the $z$ term under the reasonable assumption that the first and second nearest neighbor overlaps are more dominant.

Performing a quadratic expansion for small momenta (i.e. the ultra-low density limit discussed in the main text) we find,
\begin{align}
    \label{eq:expansion01}
     h_0(\bk) &= C_0 + \tilde{\lambda} \bk^2, \\
    h_z(\bk) &=  \frac{\lambda}{2}  (3 k_z^2 - \bk^2) \\
    h_x(\bk) &= \frac{\sqrt{3} \lambda}{2}   ( k_x^2 - k_y^2) ,    
\end{align}
where $C_0 = 3(t+u + 2v + 2w)$,  $\lambda =  \left({\frac{t - u}{2} - v + w}\right)$, and $\tilde{\lambda} = -\left( \frac{t+u}{2} + 2 (v+w) \right)$ are the collected constants referred to in the main text.

\section{Symmetry protection of band degeneracy}
The band degeneracy occurs along high-symmetry lines of the Brillouin zone including $\Gamma M$, $RX$, and $\Gamma R$. The isolated nodal points on the former two lines are protected by mirror symmetry with the mirror planes, $(110), (101), (011), (\bar{1}10),(\bar{1}01), \text{and } (0\bar{1}1)$. For example, the high-symmetry line, $\Gamma M$, lie on the $(1\bar{1}0)$ plane. The band degeneracy in this plane arises since the two bands, corresponding to the $d_{x^2 - y^2}$ and $d_{3z^2 - r^2}$ orbitals, carry the opposite $(1\bar{1}0)$-mirror eigenvalues, $\mp 1$, respectively. The presence of the mirror symmetry prevents these two bands from hybridizing with one another. 

On the other hand, the symmetry protection of the nodal line, $\Gamma R$, is due to the above mirror symmetry \emph{and} the three-fold rotation about the [111]-axis. 
On the $\Gamma R$ line, the symmetry of the Hamiltonian is given by the $C_{3v}$ point group. We have verified that the $d_{x^2 - y^2}$ and $d_{3z^2 - r^2}$ orbitals form a basis of the two-dimensional irreducible representation, $E$, of the $C_{3v}$ point group, which explains the two-fold degeneracy. 
To do this, we check that the matrix representations of the elements of the $C_{3v}$ group in the $e_g$ basis indeed have their traces identical to those of the $E$ irreducible representation.
The ferro-octupolar order, $\phi$, (or similarly, the ferro-quadrupolar order $\tilde{\phi}$) breaks all these symmetries and gaps out the two bands.

\section{$\omega$-frequency Hall current}
Using the same Boltzmann formalism as in the main text, the conductivity tensor $\sigma^{(\omega)}_{abcd}$ can be similarly shown to split into a Drude-like part and a Hall-like part, \cite{Fu2015, Moore2016, Law2023}
\begin{align}
    \sigma_{abcd}^{(\omega)D} &= \frac{3e^4}{4\hbar} \sum_n \int \td\bfk \frac{f_0 (\varepsilon_{\bfk n})\partial_a \partial_b \partial_c \partial_d \varepsilon_{\bfk n} }{(\hbar \widetilde{\omega})(\widetilde{-\omega}\hbar)(\hbar \widetilde{2\omega} )},\\
    \sigma^{(\omega) H}_{abcd} &= -\frac{e^4}{4\hbar} \left[ \frac{ \epsilon_{hab} Q_{cdh} + \epsilon_{hac} Q_{bdh} }{(\hbar\widetilde{\omega})(\widetilde{-\omega}\hbar)} + \frac{\epsilon_{had} Q_{bch} }{(\hbar \widetilde{\omega})(\hbar \widetilde{2\omega})} \right],
\end{align}
where the symmetrization is again in effect, assuming that the applied electric field is linearly polarized so that $\sigma^{(\omega)}_{abcd}$ is symmetric under the permutation of the last three indices. Like $\sigma^{(3\omega)H}_{abcd}$, $\sigma^{(\omega)H}_{abcd}$ here is also manifestly time-reversal odd and dissipationless, and its anisotropic features are quite similar to those of the $3\omega$-response in Fig. 3 in the main text.

\section{Vanishing of quadratic Hall response functions}

Consider a second-order induced current, $j^{(2)}_a(t) = \text{Re}\left[j^{(2)}_a(\omega) e^{\I \omega t}\right]$, in response to an applied electric field $E_a(t) = \text{Re}\left[E_a(\omega)e^{\I \omega t} \right]$. With the envisioned Hall-bar-like experimental setup in mind (see Fig. 1 in the main text), we once again restrict ourselves to linearly-polarized electric field, where $E_a(\omega)$ can be made real-valued by redefining the initial time. 
The induced current can be expressed in terms of a second-order response function as $j^{(2)}_a(\omega \pm \omega) \equiv \sigma_{abc}^{(\omega \pm \omega)} E_b(\omega)E_c(\pm \omega)$. 
There are two types of the induced current, $j^{(2)}_a(0)$ and $j^{(2)}_a(2\omega)$, corresponding to a rectified current and a second harmonic, respectively \cite{Fu2015}. However, due to the inversion symmetry of the system, and the inversion-odd property of $\sigma^{(\omega \pm \omega)}_{abc}$, the second-harmonic response vanishes. 

\section{Third-order Hall contribution from Berry connection polarizability}

In addition to the Berry curvature quadrupole contribution to the third-order non-linear Hall effect, there is also a contribution from the Berry connection polarizability (BCP) induced by the electric field \cite{BCP1, BCP2, BCP3}; see also Ref. \cite{BCP0, Wang2021, Liu2021} for a related contribution for the second-order non-linear transport.
The BCP, $G_{ab}^m$ for band $m$ is a rank-2 tensor given by \cite{BCP1, BCP2, BCP3},
\begin{align}
G_{ab} ^m (\bfk) = (2e)  \text{Re}\sum_{n\neq m} \frac{ (\mathcal{A}_a)_{mn} (\mathcal{A}_b)_{nm} }{\epsilon_m - \epsilon_n},
\end{align}
where $\epsilon_n$ is the Bloch eigenvalue of band $n$, and $(\mathcal{A}_a)_{mn} = \langle u_m | i \partial_{k_a} | u_n \rangle$ is the inter-band Berry connection.
One can find that under time-reversal, $G_{ab} ^m (\bfk) \xrightarrow[]{\mathcal{T}}G_{ab}^m (-\bfk)$.
Thus the BCP contribution to the conductivity tensor must be an even function of the time-reversal-odd octupolar order parameter $\phi$ (see Eq. 17 in Ref. \cite{BCP1}).
We have explicitly verified this even-in-$\phi$ property using our model. 
In this sense, our time-reversal-odd third-order Hall effect from Berry curvature quadrupole is a more suitable and more direct probe of the octupolar ordering, as the BCP can be non-vanishing even in the absence of the octupolar order.
This is in sharp contrast to the Berry curvature quadrupole contribution that is vanishing in the absence of $\phi$, due to its time-reversal-odd property.
Indeed, the non-vanishing Berry curvature quadrupole contribution can be isolated from the BCP by anti-symmetrizing the Hall signal with the field along the $\pm$[111] directions, which aligns the respective octupolar order parameter domains $\pm \phi$.

\section{Continuum Berry curvature quadrupole}

The momentum-dependent form factors present in the low-density limit of the BC quadrupole are given explicitly here:
$F_1(\mathbf{k}) = 9 k_x^8  + 12k_x^6 k_y^2  - 42k_x^4 k_y^4 +14 k_x^4k_y^2k_z^2$, where we have simplified the expression using the cubic symmetry of the integral, and $F_3(\mathbf{k}) = 4k_x^4 + 4k_y^4 - 2k_z^4 - 10k_x^2 k_y^2 -k_x^2 k_z^2 - k_y^2 k_z^2$. 

\section{Behavior of $Q_{xyz}$ for different electronic densities}

\begin{figure*}
    \centering
    \includegraphics[width = 0.7\textwidth]{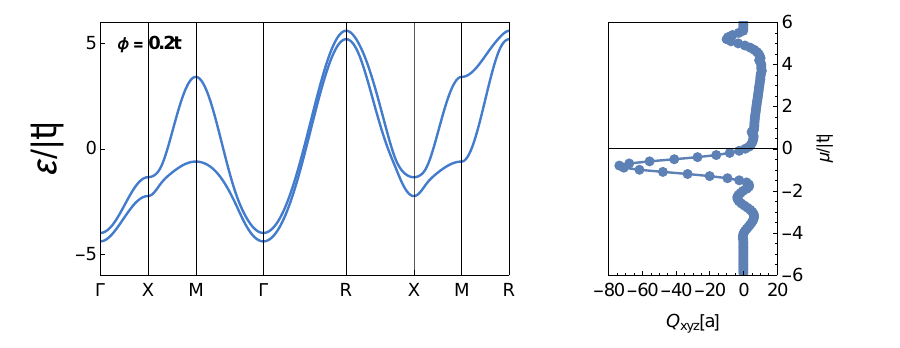}
    \caption{Dependence of the BC quadrupole $Q_{xyz}$ on the chemical potential $\mu$ for a fixed $\phi/|t| = 0.2$, plotted against the band structure.}
    \label{fig:chem}
\end{figure*}

Figure \ref{fig:chem} shows the chemical-potential dependence of $Q_{xyz}$ for a fixed $\phi/|t| = 0.2$. The large peak near $\mu = - |t|$ is attributable to the avoided band crossings such as the one along $RX$ line in the band structure Fig. 2 in the main text. 

\begin{figure}
    \centering
    \includegraphics[width=0.4\textwidth]{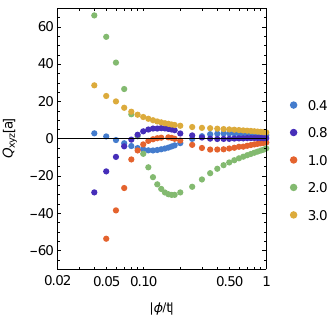}
    \caption{Dependence of the BC quadrupole on the octupolar order parameter $\phi$ at different electronic density: $(0.4/a^3, 0.8/a^3, 1.0/a^3, 2.0/a^3, 3.0/a^3)$, where $a$ is the lattice constant. By counting the spin species and the orbitals, the maximum density is $4.0/a^3$. Note that the x-axis is in logarithmic scale. Numerical results below $|\phi/t| = 0.04$ are not shown here due to a convergence issue at these higher densities.}
    \label{fig:otherDens}
\end{figure}

In Fig. \ref{fig:otherDens}, we also show the $\phi$ dependence of $Q_{xyz}$ at a few densities, other than the value in the main text. We observe that there are common features: (1) the non-monoticity and (2) the generally larger magnitude of $Q_{xyz}$ at smaller $|\phi|$.

\section{On the dependence of the BC quadrupole on the order parameter}

\begin{figure}[t]
    \centering
    \includegraphics[width=0.27\textwidth]{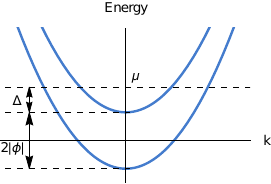}
    \caption{Schematic of the band structure near the zone center in the large-$|\phi/t|$ limit, assuming that the upper band is occupied.
    $\mu$ denotes the chemical potential which lies above the upper band's edge by an amount $\Delta$.}
    \label{fig:schematicBands}
\end{figure}

The low-electron-density limit allows a tractable analysis to shed some light on the non-monotic relation between the BC quadrupole $Q_{xyz}$ and the octupolar order parameter $\phi$. For large $|\phi/t| \gg 1$, the electrons occupy states close to the Brillouin zone's center, where the dispersion relations of the two bands are approximately given by (see Eq.\ref{eq:expansion01})
\begin{align}
\varepsilon_{\bfk, \pm} = C_0 + \tilde{\lambda} k^2 \pm |\phi|.
\label{eq_app_large_phi}
\end{align}
The density of state of the quadratic dispersions is $\rho(\xi_{\pm}) = 2\pi \sqrt{\xi_{\pm}}/\tilde{\lambda}^{3/2}$, where $\xi_{\pm} \equiv \varepsilon_{\bfk, -} - C_0 \mp |\phi|$.
We label the bands by $\pm$ just as in the main text.
For very large $|\phi/t|$, only the lower band, corresponding to $\varepsilon_{\bfk, -}$, is occupied, and subsequently, the BC quadrupole $Q_{xyz}$ scales like $1/\phi^2$, as explained in the main text. Here, we will consider the scenario when $|\phi/t|$ is further lowered so that the electrons begin to occupy the
the upper band as well, while maintaining the quadratic dispersion. 
Right after this Lifshitz transition (where the Fermi surface is now formed by both bands), we consider the case the scenario where the chemical potential lies $\Delta$ above the edge of the upper band (see Fig. \ref{fig:schematicBands}.) The total number of the electrons is given by
\begin{align}
    N &= \int_0^{2|\phi| + \Delta} \rho(\xi_-) \td \xi_- + \int_0^{\Delta} \rho(\xi_+)\td \xi_+, \nonumber\\
    &= \frac{4\pi}{3\tilde{\lambda}^{3/2}} \left[(2|\phi| + \Delta)^{3/2} + \Delta^{3/2}\right].
\end{align}
We are interested in the dependence of $Q_{xyz}$ on $\phi$, while fixing $N$ = constant. For small $\Delta \ll |\phi|$, $N \approx \frac{4\pi}{3 \tilde{\lambda}^{3/2}} \left[ (2|\phi|)^{3/2} + \frac{3 \sqrt{2|\phi|} }{2} \Delta  \right]$, so that
\begin{align}
    \Delta &\approx \frac{c_1}{\sqrt{|\phi|}} + c_2 |\phi|,
    \label{eq:relation01}
\end{align}
where $c_{1,2}$ are constants.
 Exploiting the fact that $\partial_{k_x}\partial_{k_y} \Omega_z^{\pm}$ (as seen by considering Eq. 8 in the main text in the large $\phi$ limit approximation of Eq. \ref{eq_app_large_phi})
 are approximately constant near the zone center and that they have opposite signs between the two bands at a given momentum, we obtain
\begin{align}
    Q_{xyz} & \propto \frac{\text{sgn}(\phi)}{\phi^2} \left(\int_0^{2|\phi| + \Delta} \sqrt{\xi_-} \td \xi_- - \int_0^{\Delta} \sqrt{\xi_+}\td \xi_+\right) \nonumber \\
    & = \frac{\text{sgn}(\phi)}{\phi^2} \left[\left(2|\phi| + \Delta\right)^{3/2} - \Delta^{3/2}\right], \nonumber \\
    & = \frac{3\tilde{\lambda}^{3/2} N}{4\pi} \frac{\text{sgn}(\phi)}{|\phi|} \left( 8|\phi|^2 + 12 |\phi| \Delta + 6 \Delta^2 \right).
\end{align}
Using the relation in Eq. \ref{eq:relation01}, one can show that $Q_{xyz}$ is indeed a complicated function of $\phi$, as stated in the main text. As $|\phi|$ becomes even smaller, the Fermi surfaces become non-spherical, so a straightforward estimate for the functional form of $Q_{xyz}(\phi)$ is no longer possible.

\section{Impacts of a ferro-quadrupolar ordering}

\begin{figure}
    \centering
    \includegraphics[width=0.38\textwidth]{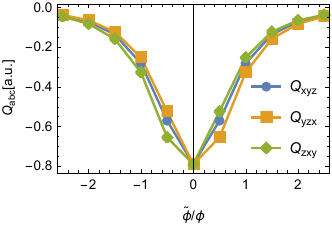}
    \caption{Impact of the ferro-quadrupolar order, $\tilde{\phi} = \langle \mathcal{O}_{22}\rangle$, in splitting the degeneracy of $Q_{xyz}, Q_{zxy}$ and $Q_{yzx}$ and in suppressing the BC quadrupoles. Here, we have chosen the model parameters as in Fig. 4(c) in the main text, with a fixed $\phi/|t| = 0.1$.}
    \label{fig:quadorder}
\end{figure}
\begin{figure}
    \centering
    \includegraphics[width=0.3\textwidth]{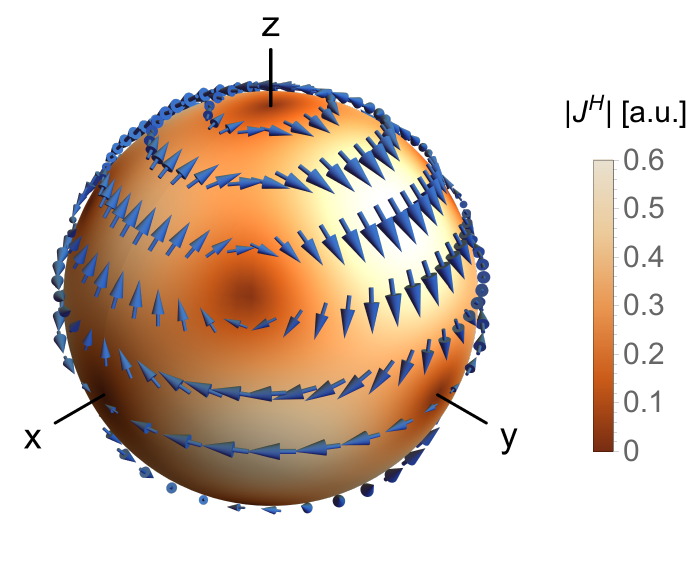}
    \caption{Anisotropy of the Hall current under the influence of coexisting quadrupolar order. 
    The induced Hall current $\bfj^H$ is depicted for $Q_{yzx}/Q_{xyz} = 1.15$ and $Q_{zxy}/Q_{xyz} = 0.75$. It is only slightly modified compared with Fig. 3 of the main text in which $Q_{yzx}/Q_{xyz} = Q_{zxy}/Q_{xyz} = 1$. 
    For instance, the center of the vortex on the [111] axis has been slightly shifted towards the x-axis.}
    \label{fig:anisotropyQuad}
\end{figure}

In this section, we consider the scenario of the system supporting a ferro-quadrupolar ordering of the type $\tilde{\phi} = \langle \mathcal{O}_{22}\rangle$ on top of the ferro-octupolar order. 
The subsequent coupling to the $e_g$ electrons is of the form, \cite{patri_kondo_2020, patri_kondo_2021},
\begin{align}
    h_{\rm Q}(\bfk) &= \tilde{\phi} \tau^x.
\end{align} 
The nonzero value of $\tilde{\phi}$ lowers the symmetry of the system from $m\bar{3}m'$ to $mmm$ magnetic point group,
which supports three independent components of $Q_{abc}$, namely $Q_{xyz}, Q_{yzx}$ and $Q_{zxy}$ (see Table III of Ref.\cite{Law2023}). These are degenerate in the absence of $\tilde{\phi}$, as illustrated in Fig. \ref{fig:quadorder}. Two main observations are found: (1) $\tilde{\phi}$ suppresses the magnitude of the BC quadrupoles, and (2) the splitting of the degeneracy is not pronounced. Feature (1) is attributed to the fact that, as $\tilde{\phi}$ increases, the BC distribution becomes increasingly smooth, which in turn leads to smaller BC derivatives. On the other hand, feature (2) leads to only a slight shift in the anisotropic properties of the induced Hall current, compared with Fig. 3 in the main text. Figure \ref{fig:anisotropyQuad} demonstrates this point: the anisotropy features of Fig. 3 are qualitatively preserved, despite the splitting, $Q_{yzx}/Q_{xyz} = 1.15$ and $Q_{zxy}/Q_{xyz} = 0.75$.

\bibliography{refs}
\end{document}